# Hall coefficient diagnostics of surface state in pressurized SmB$_6$

Yazhou Zhou[1], Priscila F. S. Rosa[2,3], Jing Guo[1], Shu Cai[1], Rong Yu[4], Sheng Jiang[5], Ke Yang[5], Aiguo Li[5], Qimiao Si[6], Qi Wu[1], Zachary Fisk[2], Liling Sun†[1,7,8]

[1]*Institute of Physics, Chinese Academy of Sciences, Beijing 100190, China*
[2]*Department of Physics and Astronomy, University of California, Irvine, CA92697, USA*
[3]*Los Alamos National Laboratory, Los Alamos, New Mexico 87545, USA*
[4]*Department of Physics, Renmin University of China, Beijing 100872, China*
[5]*Shanghai Synchrotron Radiation Facilities, Shanghai Institute of Applied Physics, Chinese Academy of Sciences, Shanghai 201204, China*
[6]*Department of Physics & Astronomy, Rice University, Houston, Texas 77005, USA*
[7]*University of Chinese Academy of Sciences, Beijing 100190, China*
[8]*Songshan Lake Materials Laboratory, Dongguan, Guangdong 523808, China*

In this study, we report the first results of the high-pressure Hall coefficient ($R_H$) measurements in the putative topological Kondo insulator SmB$_6$ up to 37 GPa. Below 10 GPa, our data reveal that $R_H(T)$ exhibits a prominent peak upon cooling below 20 K. Remarkably, the temperature at which surface conduction dominates coincides with the temperature of the peak in $R_H(T)$. The temperature dependent resistance and Hall coefficient can be well fitted by a two-channel model with contributions from the metallic surface and the thermally activated bulk states. When the bulk of SmB$_6$ becomes metallic and magnetic at ~ 10 GPa, both the $R_H(T)$ peak and the resistance plateau disappear simultaneously. Our results indicate that the $R_H(T)$ peak is a fingerprint to diagnose the presence of a metallic surface state in SmB$_6$. The high-pressure magnetic state of SmB$_6$ is robust to 180 GPa, and no evidence of superconductivity is observed in the metallic phase.

Samarium hexaboride ($SmB_6$) is a prototypical Kondo insulator in strongly correlated electron systems. At high temperatures, $SmB_6$ behaves as a correlated bad metal but undergoes a metal-to-insulator crossover upon cooling due to the hybridization between the localized *f*-electrons and the conduction electrons [1-10]. Below ~4 K, the electrical resistivity of $SmB_6$ displays a plateau, which has been a puzzling issue for decades [1,2]. This has been revealed recently to be attributed to an exotic metallic surface state that coexists with a bulk insulating state [6-9,11-17]. Thus, $SmB_6$ could be the first example of a new class of topological insulators with strong electronic correlations [18-20]. Considerable experimental efforts have been made to confirm the topological nature of the surface states in $SmB_6$, and the correlation between the resistance plateau (RP) and a metallic surface state was established [6-9,11-17]. Importantly, a two-channel conductivity model was found to describe well the temperature dependent resistance and Hall coefficient of $SmB_6$ by tuning sample thickness and gate voltage [21,22]. Pressure is a clean and effective way of tuning interactions in solids with multiple degrees of freedom without introducing chemical complexity. Therefore, pressure has been successfully adopted in the studies of a broad variety of materials [23-33]. Although $SmB_6$ at high pressures has been extensively investigated before the discovery of its metallic surface state [32-40], including measurements of Hall coefficient and resistivity under pressures to 8 GPa [41], high-pressure Hall coefficient studies of the intimate correlations among the exotic surface and bulk states, crystal structure and correlated electrons are still lacking. In this work, we are the first to perform the high-pressure measurements on the high quality $SmB_6$ single crystals to identify how the metallic surface and insulating bulk states evolve under pressure from a perspective of $R_H(T)$.

In Fig.1a we show the temperature dependence of electrical resistance measured in a $SmB_6$ sample upon cooling under pressures up to 36.8 GPa. The resistance of the sample at 1.1 GPa displays a continuous increase upon cooling and then exhibits a plateau below 5 K, identical to the behavior at ambient pressure [2,4,12,42-46]. Further, the resistance plateau is clearly visible at pressures below 7.6 GPa, but becomes indistinguishable at 8.5 GPa. Further compression to pressures above 9.7 GPa leads to the disappearance of the resistance plateau and to a substantial drop of the magnitude of the electrical resistance at low temperature, a signature of an insulator-to-metal transition. It is worth noting that the critical pressure of the insulator-to-metal transition found in this study is ~ 10 GPa (see the experimental details in Supplementary Information), in excellent agreement with the critical pressure applied to the sample through a gas transmitting medium [38].

High-pressure Hall coefficient measurements were performed upon warming after the resistance measurements. As shown in Fig.1b, in the temperature range of 1.8 K - 30 K, $R_H(T)$ is negative at all pressures, indicating that the electron carriers are dominant. Remarkably, the plot of $R_H$ versus temperature displays a dome-like behavior in the temperature range below 10.7 GPa (Fig.1b and inset). The dome-like $R_H(T)$ observed in the $SmB_6$ sample is attributed to the combined contribution from surface and the bulk states. Above 10 K, the insulating behavior of the bulk state is significantly dominant (Fig.1a), which leads to an increase in $R_H(T)$, while, when the metallic surface state sets in at the temperature below the formation of the resistance plateau, $R_H$ shows a decease upon cooling due to the dominance of the metallic surface state [21]. The temperature dependence of both resistance and Hall coefficient obtained under pressure can be well fitted by a two-channel model consisting of a thermally activated bulk in parallel with a temperature-independent surface state [21,22]. The resistance can be described as:

$$R = \left(1/R_S + 1/R_B\right)^{-1} \tag{1}$$

where $R_S$ and $R_B$ represent the resistance from surface and bulk channels, respectively. Here, $R_B = R_{B_0} e^{E_g/k_B T}$, in which $E_g$ is the activated energy gap, $k_B$ is Boltzmann constant, $R_{B_0}$ is the bulk resistance in the high-temperature limit and $T$ is temperature.

The Hall coefficient can be expressed as:

$$R_H = \left(R_{H-S} d\rho_B^2 + R_{H-B}(d\rho_S)^2\right) \Big/ (d\rho_S + \rho_B)^2 \tag{2}$$

where $R_{H-S}$ and $R_{H-B}$ are the Hall coefficient of the surface and bulk states, respectively. $\rho_S$ and $\rho_B$ are resistivity the surface and bulk states, $d$ is the thickness of the sample (the estimated $d$ used in the fit was ~ 10 μm).

The solid lines in Fig.1a and 1b are the fit results. At low pressures, our results are well described by the two-channel model. The surface state dominates the electrical conductivity when the resistance plateau appears, whereas the bulk insulating state dominates above the temperature of the resistance plateau formation. At ~ 10.7 GPa, $R_H(T)$ is featureless, indicating that the conductance of the bulk state is comparable with that of the surface state due to the pressure-induced metallization. Our results demonstrate that the dome-like $R_H(T)$ can be taken as a fingerprint to distinguish the coexistence of the metallic surface and insulating bulk states in $SmB_6$ or other topological insulators.

Magnetic order has been found previously in pressurized $SmB_6$ by nuclear forward scattering of synchrotron radiation measurements [47]. The magnetic ordering temperature ($T_M$) has been confirmed to lie in the 10 -12 K range at ~ 10 GPa [47]. We find that the mid-point temperatures ($T'$) of the resistance drop in metallic $SmB_6$ are close to its corresponding $T_M$ measured by nuclear magnetic resonance measurements. Note that the feature of $T'$ in the resistance curve persists up to 180 GPa (as indicate by arrows in Fig.2). If $T'$ is taken as a characteristic temperature of $T_M$ (Fig.2a), it is surprising to find that $T_M$ is robust under pressures as high as 180 GPa (Fig.2b). Because $R(T)$ curves measured in the pressure range of 10 GPa - 180 GPa exhibit similar behavior and no structure phase transition is observed under pressures to 167 GPa (Fig.3), we propose that the magnetically ordered state remains in the pressurized metallic phase throughout this pressure range.

Remarkably, $R(T)$ shows a distinguishable feature of the magnetic order state at 7.6 GPa and 8.5 GPa, whereas $R_H(T)$ still displays a peak behavior, indicating that $SmB_6$ is in an intermediate state that bridges the high-pressure metallic magnetic state and the low-pressure state with coexisting metallic surface and insulating bulk. This intermediate state deserves further theoretical and experimental investigations.

We summarize our results in Fig.4. $SmB_6$ hosts a metallic surface state below 10 GPa, which is characterized by the resistance plateau and the peak in $R_H(T)$. This peak decrease slightly with increasing pressure and eventually disappear at ~ 10 GPa as the bulk state of the sample becomes metallic state (Fig.4a). We also find that the mid-point of the resistance drop, $T'(P)$, coincides with the magnetic transition temperature $T_M$ detected by nuclear scattering forward measurements [47], and it is present to 180 GPa. These results suggest that a robust magnetically ordered state is stabilized, which prevents the emergence of superconductivity.

The corresponding pressure dependence of the $R_H$ obtained at 1.8 K is shown in Fig. 4b. Below ~10 GPa, $R_H$ decreases with increasing pressure and stays almost constant in the metallic magnetic state. It is known that $SmB_6$ is a mixed valence compound at ambient pressure with valence $v_{Sm}$ ~ 2.6. The application of pressure drives the valence change of Sm ions from delocalized to localized state, *i.e.*, the concentration of magnetic $Sm^{3+}$ ions is enhanced upon compression. Previous high-pressure absorption measurements [33,48-50] indicated that its mean valence is very close to 3+ at P>~10 GPa. The pressure-induced valence change of $Sm^{2+} \rightarrow$ ($Sm^{3+}+5d$), together with its stable cubic lattice structure, should be responsible for the robustness of long-ranged magnetic order [35,48-50].

In conclusion, the coexistence of surface and bulk states and their evolution with pressure in a putative topological insulator $SmB_6$ has been revealed by the prominent feature of the temperature dependent Hall coefficient for the first time. The intimate correlation between the low-temperature $R_H$ and the exotic surface state suggests that $R_H(T)$ is one of the most useful diagnostic methods to identify the existence of the exotic surface state in $SmB_6$ and other topological insulators. Furthermore, we find the extraordinary robustness of the crystal structure and metallic state in compressed

SmB$_6$ up to 180 GPa and no superconductivity is observed in the pressure range investigated.

**Acknowledgements**

The authors are indebted to P. Coleman, J. Thompson, L. Greene for stimulating discussions. The work in China was supported by the National Key Research and Development Program of China (Grant No., 2016YFA0300300, 2017YFA0302900 and 2017YFA0303103), the NSF of China (Grants No. 91321207, No. 11427805, No. U1532267, No. 11604376, 11374361, No. 11522435), the Strategic Priority Research Program (B) of the Chinese Academy of Sciences (Grant No. XDB07020300). Work at Los Alamos was performed under the auspices of the U.S. Department of Energy, Division of Materials Sciences and Engineering. Work at Rice University was supported by the ARO Grant No. W911NF-14-1-0525 and the Robert A. Welch Foundation Grant No. C-1411.

†Corresponding authors

llsun@iphy.ac.cn

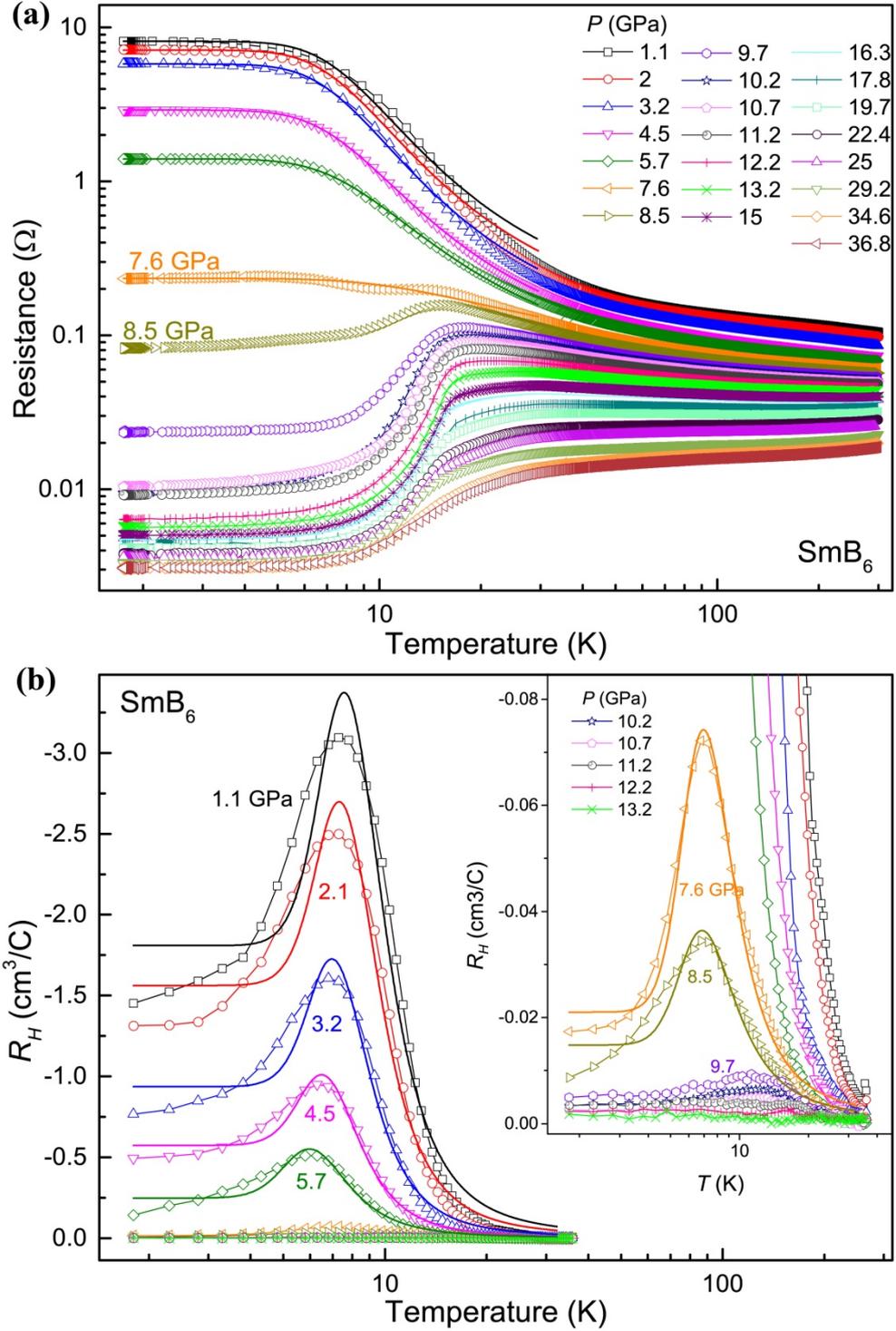

**Figure 1 Transport properties in pressurized SmB$_6$.** (a) Temperature dependence of resistance from 1.7 to 300 K at different pressures in log-log scale. Solid line is fitting by two-channel model. (b) Hall coefficient ($R_H$) as a function of temperature

measured at different pressures. The inset is the large view of $R_H(T)$ obtained at higher pressure. Solid lines are fitting result.

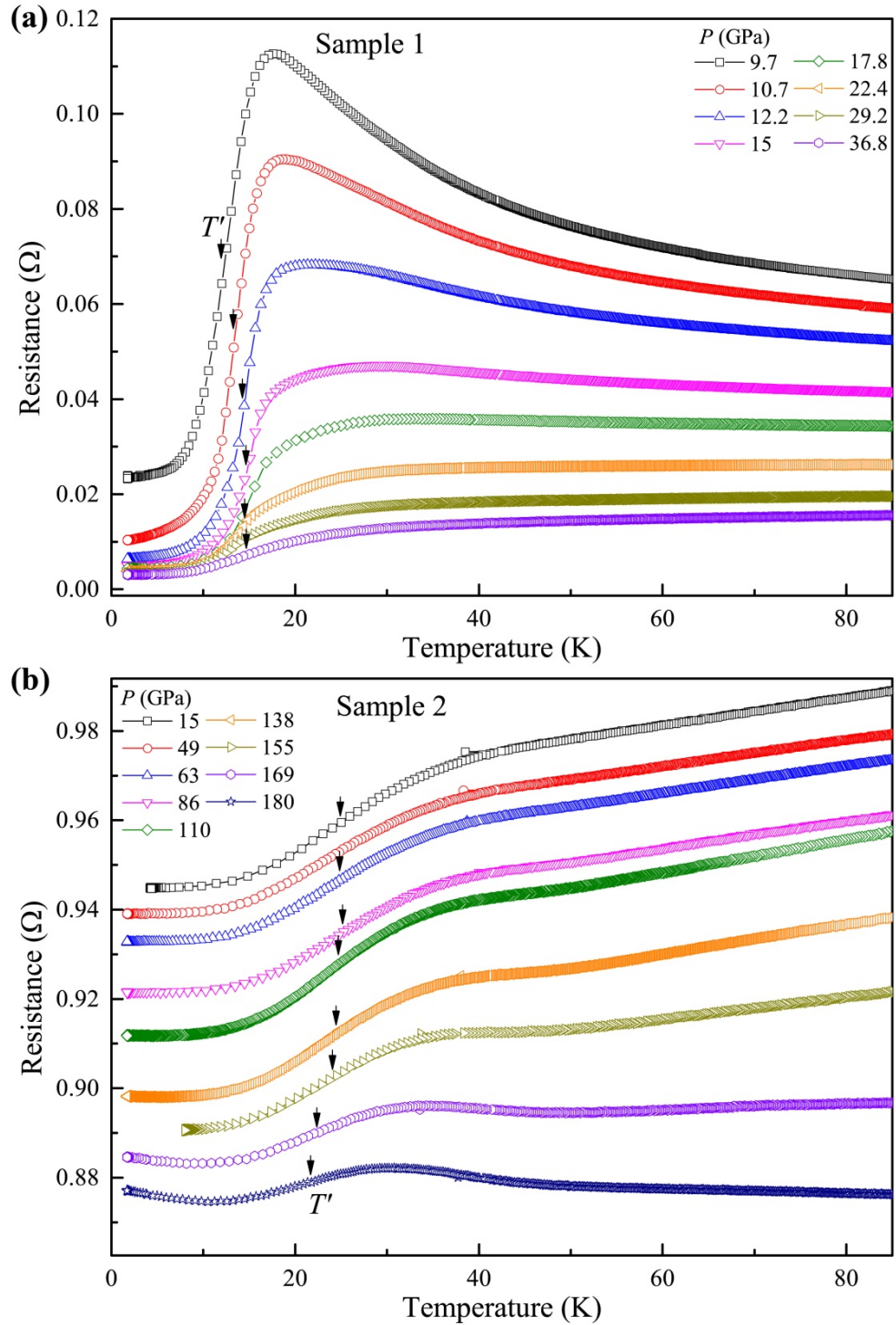

**Figure 2 High pressure behavior of metallic SmB$_6$.** (A) Resistance versus temperature in the pressure range of 9.7 ~ 36.8 GPa for Sample 1. (B) Temperature dependence of resistance under pressure up to 180 GPa for Sample 2. *T″* represents maximum of *dR/dT* curves indicating the resistance drop in the metallic phase.

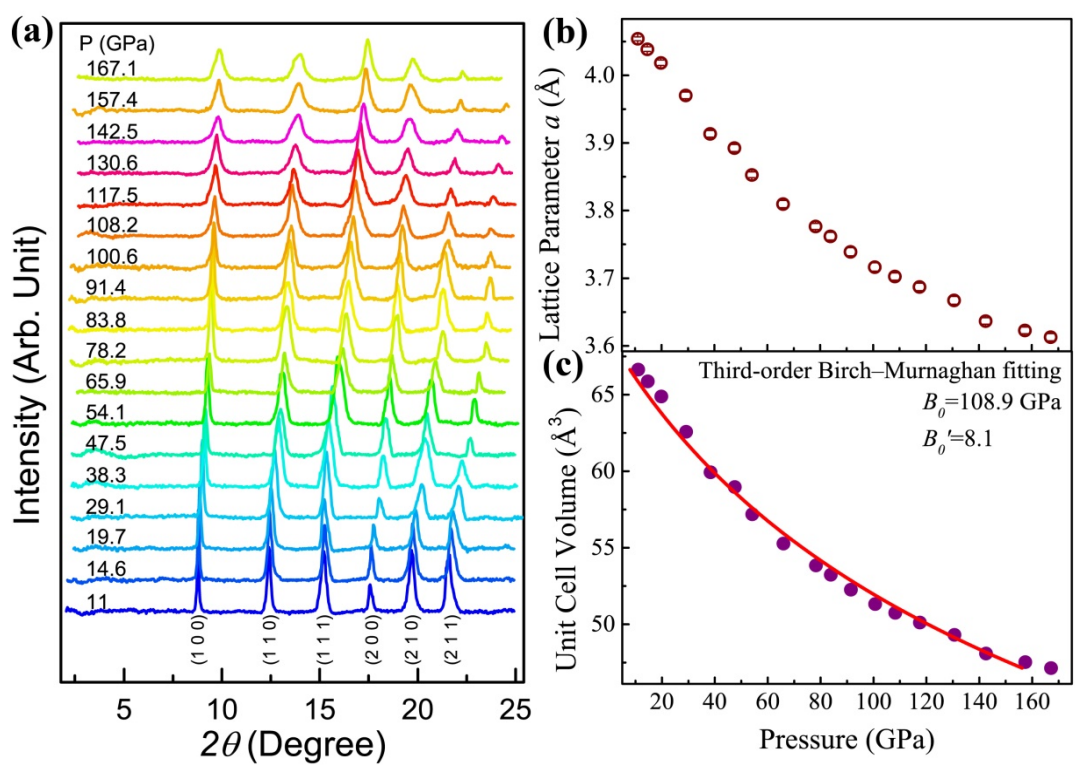

**Figure 3 High pressure structure information of SmB$_6$.** (a) X-ray diffraction patterns of SmB$_6$ collected at different pressures. (b) and (c) Pressure dependences of lattice parameter and volume.

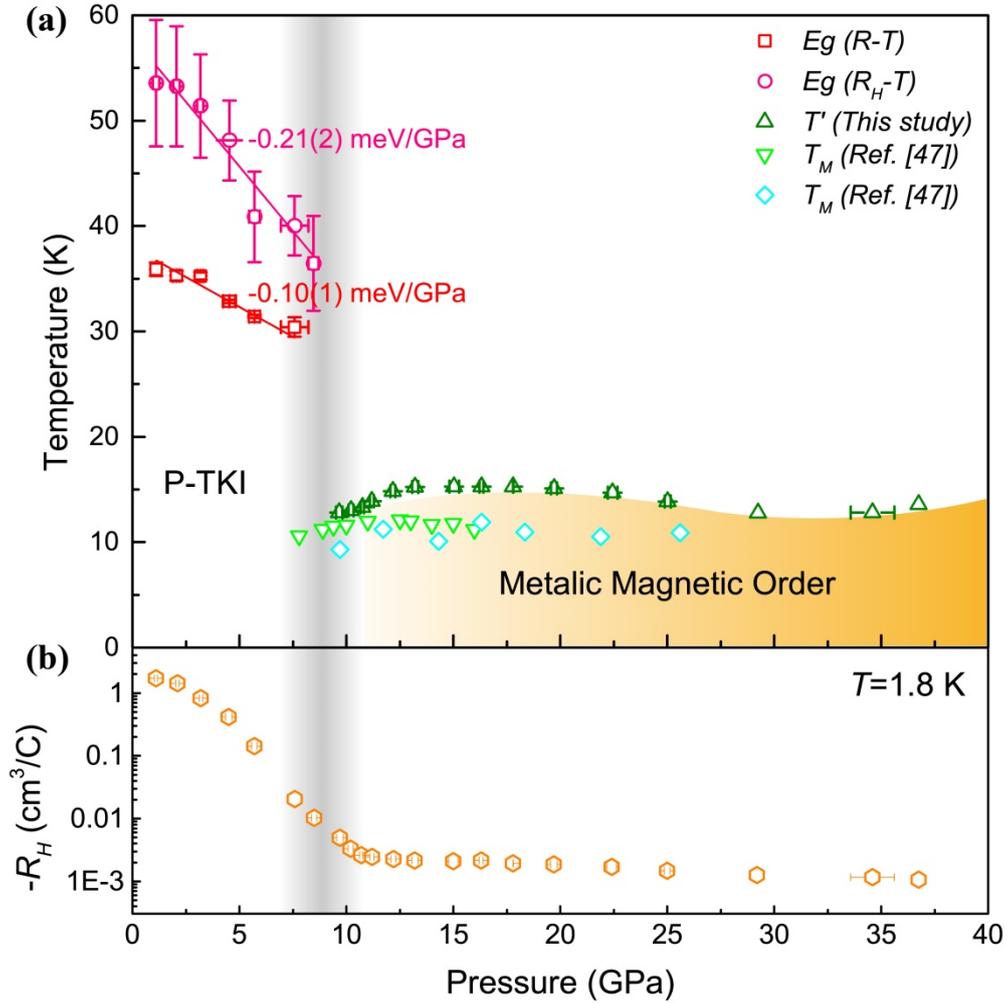

**Figure 4 Pressure-temperature phase diagram and Hall coefficient at 1.8 K**. (a) Plot of pressure versus characteristic temperatures. P-TKI stands for putative topological Kondo insulator. Pink circle and red square are the activated gap $E_g$ obtained from temperature-dependent $R$-$T$ curves and Hall coefficient respectively and is converted to temperature by equation of (1) and (2). Olive triangle is characteristic temperatures $T'$ obtained from our $R$-$T$ curves. Green inverted triangle and cyan rhombus is magnetic ordering temperature taken from Ref. [51]. (b) Plot of Hall coefficient ($R_H$) versus pressure obtained at 1.8 K. Gray interval indicates the pressure region of insulator-to-metal transition.